%% file: main.tex
\documentclass[aps,prl,twocolumn,superscriptaddress,longbibliography,amsmath,amssymb]{revtex4-2}
\usepackage{xcolor}
\usepackage{graphicx}

\usepackage[separate-uncertainty = true,multi-part-units=single]{siunitx}
\usepackage[version=4,arrows=pgf-filled,
textfontname=sffamily,
mathfontname=mathsf]{mhchem}

\usepackage[utf8]{inputenc}
\usepackage{amssymb,amsmath}
\usepackage{tabularx}
\usepackage{multirow}
\usepackage{booktabs}
\usepackage{colortbl}
\usepackage{tabularx}
\usepackage{color}
\usepackage{systeme}
\usepackage{hyperref}
\usepackage{cleveref}

\usepackage[bbgreekl]{mathbbol}
\usepackage{mathrsfs}
\usepackage[scr=esstix,cal=boondox]{mathalfa} 

\usepackage{multibib}
\newcites{supp}{Supplementary references}

\newcommand{\vect}[1]{\boldsymbol{#1}}
\renewcommand{\Re}{\mathop{\rm Re}}
\renewcommand{\Im}{\mathop{\rm Im}}

\begin{document}
\title{Nonradiative energy transfer between boron vacancies in hexagonal boron nitride and other 2D materials}

\input{Sections/00a-authors}
\begin{abstract}
\input{Sections/00b-abstract}
\end{abstract}

\maketitle 

\input{Sections/01-intro}

\input{Sections/02-in-this-letter}
\input{Sections/03-experimental-details}
\input{Sections/04-results}
\input{Sections/05-discussion}

\input{Sections/06-conclusion}
\input{Sections/07-acknowledgements}

\bibliographystyle{apsrev4-2}
\bibliography{references2}


\clearpage
\newpage
\onecolumngrid

\setcounter{figure}{0}
\renewcommand{\figurename}{Fig.}
\renewcommand{\thefigure}{S\arabic{figure}}

\setcounter{table}{0}
\renewcommand{\tablename}{Table}
\renewcommand{\thetable}{S\arabic{table}}

\setcounter{page}{1}
\renewcommand{\thetable}{S\arabic{page}}

\setcounter{equation}{0}
\renewcommand{\theequation}{S\arabic{equation}}


\renewcommand{\figurename}{Figure}
\def\dfrac{\displaystyle\frac}
\newcommand{\Sp}{\mathop{\rm Sp}\nolimits}
\newcommand{\St}{\mathop{\rm St}\nolimits}
\newcommand{\erf}{\mathop{\rm erf}\nolimits}
\newcommand{\arch}{\mathop{\rm Arch}\nolimits}
\newcommand{\floor}{\mathop{\rm floor}\nolimits}
\newcommand{\rot}{\mathop{\rm rot}\nolimits}
\newcommand{\divv}{\mathop{\rm div}\nolimits}
\renewcommand{\Re}{\mathop{\rm Re}}
\renewcommand{\Im}{\mathop{\rm Im}}
\newcommand{\Ei}{\mathop{\rm Ei}}
\renewcommand {\i}{{\rm i}}

\newcommand{\add}[1]{\textcolor{magenta}{#1}}
\newcommand{\commentMisha}[1]{\textcolor{magenta}{{\it #1}}}

\renewcommand{\baselinestretch}{1.33} 
\renewcommand{\thepage}{S\arabic{page}}
\renewcommand{\thetable}{S\arabic{table}}
\renewcommand{\thefigure}{S\arabic{figure}}
\renewcommand{\theequation}{S\arabic{equation}}

\begin{center}{\vspace{-2cm} {\large Supplemental Material for}\\ \vspace{0.5cm}
\textbf {\Large Non-radiative energy transfer between boron vacancies in hexagonal boron nitride and other 2D materials}
}



\vspace{1cm}

Jules Frauni\'{e},$^{1}$ Mikhail M. Glazov,$^{2}$ S\'{e}bastien Roux,$^{1}$ Abraao Cefas Torres-Dias,$^{1}$ Cora Crunteanu-Stanescu,$^{1}$ Tom Fournier,$^{1}$ Maryam S. Dehaghani,$^{1}$ Tristan Clua-Provost,$^{3}$ Delphine Lagarde,$^{1}$ Laurent Lombez,$^{1}$ Xavier Marie,$^{1,4}$ Benjamin Lassagne,$^{1}$ Thomas Poirier,$^{5}$ James H. Edgar,$^{5}$ Vincent Jacques,$^{3}$ and Cedric Robert$^{1}$

\vspace{0.5em}

\small
\textit{$^1$Universit\'{e} de Toulouse, INSA-CNRS-UPS, LPCNO, 135 Av. Rangueil, 31077 Toulouse, France\\}
\textit{$^2$Ioffe Institute, Russian Academy of Sciences, 194021 St. Petersburg, Russia\\}
\textit{$^3$Laboratoire Charles Coulomb, Universit\'e de Montpellier and CNRS, 34095 Montpellier, France\\}
\textit{$^4$Institut Universitaire de France, 75231, Paris, France\\}
\textit{$^5$Tim Taylor Department of Chemical Engineering, Kansas State University, Manhattan, Kansas 66506, USA}

\vspace{2em}
\end{center}


\input{SectionSupp/01_Sample_fabrication}

\input{SectionSupp/02_Experimental_details}
\input{SectionSupp/03_Additional_data}
\input{SectionSupp/04_Model_test}



\end{document}

%% file: Sections/00a-authors.tex
\author{Jules Frauni\'e}
\affiliation{Universit\'e de Toulouse, INSA-CNRS-UPS, LPCNO, 135 Av. Rangueil, 31077 Toulouse, France}
\author{Mikhail M. Glazov}
\affiliation{Ioffe Institute, Russian Academy of Sciences, 194021 St. Petersburg, Russia}
\author{S\'ebastien Roux}
\affiliation{Universit\'e de Toulouse, INSA-CNRS-UPS, LPCNO, 135 Av. Rangueil, 31077 Toulouse, France}
\author{Abraao Cefas Torres-Dias}
\affiliation{Universit\'e de Toulouse, INSA-CNRS-UPS, LPCNO, 135 Av. Rangueil, 31077 Toulouse, France}
\author{Cora Crunteanu-Stanescu}
\affiliation{Universit\'e de Toulouse, INSA-CNRS-UPS, LPCNO, 135 Av. Rangueil, 31077 Toulouse, France}
\author{Tom Fournier}
\affiliation{Universit\'e de Toulouse, INSA-CNRS-UPS, LPCNO, 135 Av. Rangueil, 31077 Toulouse, France}
\author{Maryam S. Dehaghani}
\affiliation{Universit\'e de Toulouse, INSA-CNRS-UPS, LPCNO, 135 Av. Rangueil, 31077 Toulouse, France}
\author{Tristan Clua-Provost}
\affiliation{Laboratoire Charles Coulomb, Universit\'e de Montpellier and CNRS, 34095 Montpellier, France}
\author{Delphine Lagarde}
\affiliation{Universit\'e de Toulouse, INSA-CNRS-UPS, LPCNO, 135 Av. Rangueil, 31077 Toulouse, France}
\author{Laurent Lombez}
\affiliation{Universit\'e de Toulouse, INSA-CNRS-UPS, LPCNO, 135 Av. Rangueil, 31077 Toulouse, France}
\author{Xavier Marie}
\affiliation{Universit\'e de Toulouse, INSA-CNRS-UPS, LPCNO, 135 Av. Rangueil, 31077 Toulouse, France}
\affiliation{Institut Universitaire de France, 75231, Paris, France}
\author{Thomas Blon}
\affiliation{Universit\'e de Toulouse, INSA-CNRS-UPS, LPCNO, 135 Av. Rangueil, 31077 Toulouse, France}
\author{Benjamin Lassagne}
\affiliation{Universit\'e de Toulouse, INSA-CNRS-UPS, LPCNO, 135 Av. Rangueil, 31077 Toulouse, France}
\author{Thomas Poirier}
\affiliation{Tim Taylor Department of Chemical Engineering, Kansas State University, Manhattan, Kansas 66506, USA}
\author{James H. Edgar}
\affiliation{Tim Taylor Department of Chemical Engineering, Kansas State University, Manhattan, Kansas 66506, USA}
\author{Vincent Jacques}
\thanks{Corresponding author: vincent.jacques@umontpellier.fr, cerobert@insa-toulouse.fr}
\affiliation{Laboratoire Charles Coulomb, Universit\'e de Montpellier and CNRS, 34095 Montpellier, France}
\author{Cedric Robert}
\thanks{Corresponding author: vincent.jacques@umontpellier.fr, cerobert@insa-toulouse.fr}
\affiliation{Universit\'e de Toulouse, INSA-CNRS-UPS, LPCNO, 135 Av. Rangueil, 31077 Toulouse, France}

%% file: Sections/00b-abstract.tex
Boron vacancies ($V_{B}^{-}$) in hexagonal boron nitride (hBN) have emerged as a promising platform for two-dimensional quantum sensors capable of operating at atomic-scale proximity. However, the mechanisms responsible for photoluminescence quenching in thin hBN sensing layers when placed in contact with absorptive materials remain largely unexplored. In this Letter, we investigate non-radiative Förster resonance energy transfer (FRET) between $V_{B}^{-}$ centers and either monolayer graphene or 2D semiconductors. Strikingly, we find that the FRET rate is negligible for hBN sensing layers thicker than 3 nm, highlighting the potential of $V_{B}^{-}$ centers for integration into ultra-thin quantum sensors within van der Waals heterostructures. Furthermore, we experimentally extract the intrinsic radiative decay rate of $V_{B}^{-}$ defects.

%% file: Sections/01-intro.tex
In recent years, quantum sensing using optically active spin defects in two-dimensional (2D) materials has attracted growing interest \cite{vaidya_quantum_2023}. Unlike spin-based sensors embedded in three-dimensional host matrices, such as nitrogen-vacancy (NV) centers in diamond, 2D platforms promise atomic-scale proximity between the spin sensor and the target sample. This extreme proximity is particularly advantageous for detecting weak magnetic-field sources, as the magnetic field produced by a localized source decays as inverse cube of the sensor-to-sample distance.
It may also be critical for applications like nuclear magnetic resonance (NMR) detection and could extend 2D quantum metrology to diverse scientific domains, including low-dimensional superconductors, quantum materials, and complex chemical and biological environments. Furthermore, 2D sensing platforms allow seamless integration into van der Waals (vdW) heterostructures, paving the way for exploring novel quantum phenomena occurring at the interfaces between different 2D materials.

Hexagonal boron nitride (hBN) is an attractive material for 2D quantum sensors due to its wide band gap and ability to host optically stable emitters from the ultraviolet to the near-infrared \cite{aharonovich_quantum_2022}. Among these, the negatively charged $V_{B}^{-}$ center has emerged as the most promising defect for quantum sensing \cite{gottscholl_initialization_2020}. It emits a broad PL spectrum around 820 nm, stable down to few-layer thicknesses and at room temperature \cite{durand_optically_2023}.
Like the NV center, $V_{B}^{-}$ has a spin-triplet ground state (S=1), whose electron spin resonance frequencies can be interrogated by optical means \cite{gottscholl_spin_2021}. Proof-of-concept demonstrations have shown magnetic imaging, temperature sensing, and strain mapping \cite{kumar_pawan_magnetic_2022, zhou_sensing_2024, healey_quantum_2023, huang_wide_2022}, though mostly in thick hBN flakes ($>$10 nm), far from the regime of 2D quantum sensing.

Moreover, the $V_{B}^{-}$ defect is known to be a dim emitter, which currently limits its detection to ensembles and raises questions about the feasibility of quantum sensing with ultra-thin hBN layers ($<$5-10 nm). The origin of this weak light emission has recently been the subject of intense debate in the community. Theoretical predictions suggest an intrinsic radiative decay rate as low as $\sim 10^{5}\,\mathrm{s^{-1}}$ \cite{ivady_ab_2020, reimers_photoluminescence_2020}. In contrast, some experimental photodynamics studies have reported radiative rates four orders of magnitude higher ($\sim 10^{9}\,\mathrm{s^{-1}}$), based on multi-parameter fitting models, and attribute the dim emission to strong non-radiative effects \cite{whitefield_magnetic_2025, lee_intrinsic_2025}. Thus, an unambiguous determination of this radiative rate remains elusive, despite its critical role in evaluating the potential sensitivity of $V_{B}^{-}$ based ultra-thin sensor.
	
Achieving quantum sensing at the atomic scale also requires verifying that the optical properties of the spin defects remain intact when an ultra-thin sensing layer is placed directly on the sample. Indeed, a reduction of photoluminescence (PL) in such a configuration can arise from three main mechanisms:

\textit{Optical effects} – Modifications of absorption or collection efficiency due to the surrounding dielectric environment \cite{gerard_quantum_2024}. Depending on layer thicknesses or reflective/absorptive interfaces, PL may be quenched or enhanced, an effect that can be optimized through cavity design and transfer matrix modeling.

\textit{Charge-state conversion} – Spin defects may switch between radiative and non-radiative charge states. For instance, the charge state of NV centers in diamond can be modified near surfaces as a result of band bending, which limits their sensing capabilities \cite{santori_vertical_2009, hauf_chemical_2011, grotz_charge_2012}. External gating can partially stabilize the desired charge state \cite{yu_electrical_2022, white_electrical_2022, steiner_current-induced_2025}. We showed in our previous work that most of boron vacancies in hBN are intrinsically in their optically active charge state which exclude this mechanism as the main source of quenching for this particular platform \cite{fraunie_charge_2025}.

\textit{Non-radiative energy transfer (NRET)} – Excited-state energy can be transferred to nearby absorptive layers such as metals, semimetals, or semiconductors via Förster and Dexter mechanisms \cite{basko_energy_2000, kos_different_2005, swathi_resonance_2008, federspiel_distance_2015}. Understanding the efficiency of NRET between spin defects embedded in ultra-thin layers of 2D materials and absorptive layers is essential for advancing nanoscale quantum sensing in vdW heterostructures. This is especially relevant given that most 2D magnets are either metallic (CrTe$_2$, Fe$_x$GeTe$_2$ ...) or semiconducting (CrX$_3$, CrSBr ...). It is also essential to consider the possibility of integrating 2D quantum sensors with 2D superconductors or moiré-induced quantum phases, such as those in twisted graphene or transition metal dichalcogenides (TMD).

%% file: Sections/02-in-this-letter.tex
In this Letter, we investigate the efficiency of non-radiative energy transfer (NRET) between $V_{B}^{-}$ emitters in hBN and either monolayer graphene (MLG) or a TMD flake as archetypal 2D semimetal and 2D semiconductor materials. This study enables us to directly measure the experimental value of the intrinsic radiative rate consitent with the theoretical calculation. We also predict the efficiency of NRET for any 0D emitters in hBN depending on their intrinsic quantum yield and emission wavelength.

NRET can occur through two distinct short-range mechanisms: Förster \cite{forster_zwischenmolekulare_1948} and Dexter \cite{dexter_theory_1953} transfer. In both cases, the energy of an excited donor—here, the $V_{B}^{-}$ center—is transferred to a nearby acceptor (MLG or the 2D semiconductor) without photon emission.

The Dexter mechanism involves direct electron exchange between donor and acceptor orbitals. Its efficiency depends on wavefunction overlap, which for graphene is strongly confined to the atomic plane. As a result, Dexter transfer is only relevant at extremely short distances ($<$10 Å), and its coupling strength is typically four orders of magnitude weaker than Förster for zero-dimensional emitters on graphene \cite{malic_forster-induced_2014}. We therefore neglect it in the following.

\begin{figure}[h!]
\label{Fig1}
\includegraphics[width=0.7\linewidth]{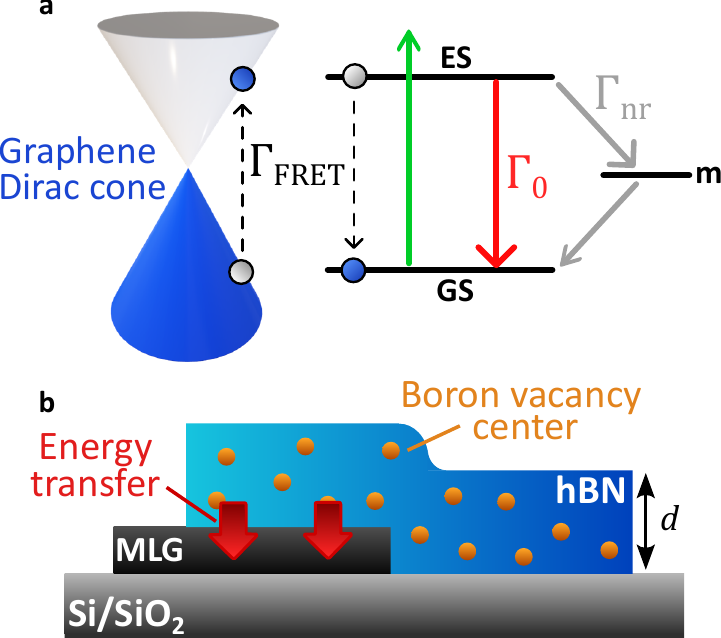}
\caption{a) Sketch of the decay rates of a $V_{B}^{-}$ center close to graphene. ES and GS are the excited and ground states of the defect, $\Gamma_{0}$ is the radiative decay rate, $\Gamma_{\mathrm{nr}}$ is the relaxation rate to the metastable state m and $\Gamma_{\mathrm{FRET}}$ the FRET rate between the emitter and graphene. b) Sketch of the samples.}
\label{fig1}
\end{figure}

Förster resonance energy transfer (FRET) arises from dipole-dipole interactions, where the donor transfers excitation energy to the acceptor via a virtual photon (see Fig. \ref{fig1}a for the case of $V_{B}^{-}$ near MLG). Graphene is known to be an exceptionally efficient FRET acceptor owing to three properties \cite{gaudreau_universal_2013, federspiel_distance_2015}: (i) its gapless band structure, which enables coupling with the radiation across a wide range of emitters; (ii) its strictly two-dimensional character, leading to a characteristic $z_0^{-4}$ dependence of the FRET rate (with $z_0$ the emitter–graphene separation); (iii) its linear band dispersion, which provides a high density of states directly enhancing the coupling strength.

FRET between 0D emitters and graphene has been extensively studied in other contexts \cite{koppens_graphene_2011, chen_energy_2010}. For instance, Gaudreau \textit{et al.}\cite{gaudreau_universal_2013} observed quenching and modified non-radiative decay rates for molecular emitters up to 15 nm from MLG , while Tisler \textit{et al.} \cite{tisler_single_2013} reported strong quenching of a single NV center even at separations of $\sim$30 nm . These results suggest that FRET is likely to play a major role in the non-radiative decay of color centers embedded in moderately thin hBN flakes deposited on graphene.

%% file: Sections/03-experimental-details.tex
A schematic of the hBN/graphene structures examined in this study is shown in Fig. \ref{fig1}b. A thin hBN flake of thickness $d$ with an homogeneous density of $V_{B}^{-}$ centers is exfoliated and deposited partially on a MLG and onto a SiO$_2$/Si substrate. Continuous-wave (cw) and time-resolved photoluminescence (TRPL) is collected at room temperature. Details about the sample fabrication and experimental setups are given in the Supplemental Material \cite{supplementary-material}.

%% file: Sections/04-results.tex
\begin{figure*}[t!]
\label{Fig2}
\includegraphics[width=0.85\linewidth]{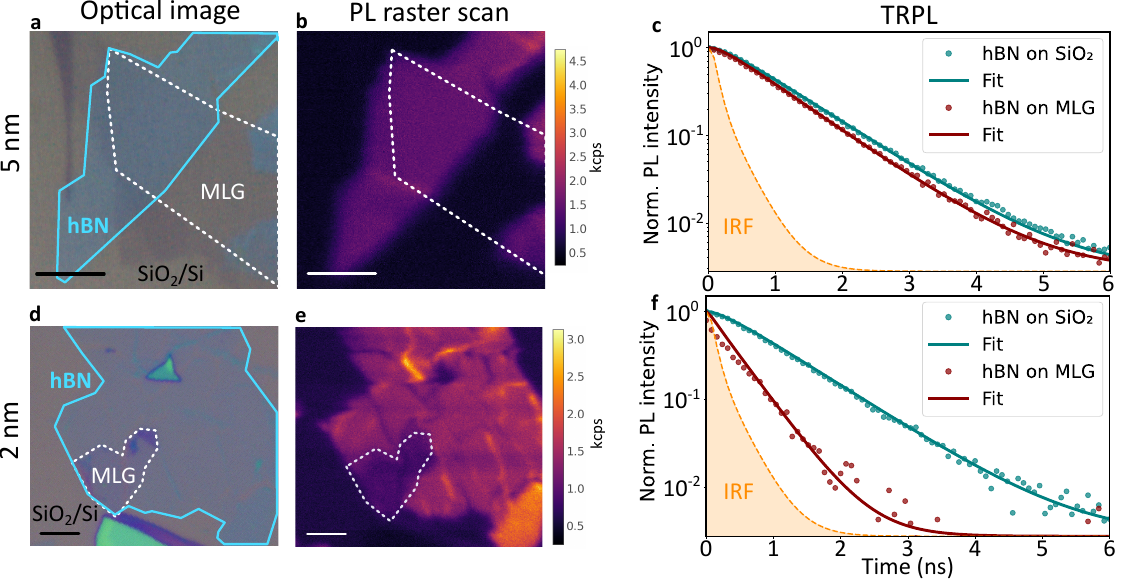}
\caption{a,d) Optical image of a 5 and 2 nm hBN flake partially deposited on MLG and SiO$_2$/Si. b,e) Corresponding cw PL raster scans. c,f) TRPL data and fits.}
\label{fig2}
\end{figure*}

Figures~\ref{fig2}a and \ref{fig2}b show the optical image and PL raster scan of a 5 nm hBN flake partially covering MLG and SiO$_2$/Si. Remarkably, the PL count rate is nearly identical on both regions, indicating that FRET-mediated non-radiative decay is not significant in this case. This conclusion is supported by the TRPL data in Fig.~\ref{fig2}c; the lifetime for the flake on MLG (red points) is only a marginally shorter.  

This behavior can be qualitatively explained by the simple three level model of Fig.~\ref{fig1}a. An excited $V_{B}^{-}$ center can decay radiatively , non-radiatively through FRET, or non-radiatively to a metastable state at rates $\Gamma_{0}$, $\Gamma_{\mathrm{FRET}}$ and $\Gamma_{\mathrm{nr}}$ respectively. The total decay rate is thus:
\begin{equation}
\Gamma_{\mathrm{tot}} = \Gamma_{0} + \Gamma_{\mathrm{nr}} + \Gamma_{\mathrm{FRET}}
\end{equation}
and the PL intensity scales as
\begin{equation}
I_{\mathrm{PL}} \propto \frac{\Gamma_{0}}{\Gamma_{0} + \Gamma_{\mathrm{nr}} + \Gamma_{\mathrm{FRET}}}
\end{equation}

For 0D emitters at a distance $z_0$ from graphene, $\Gamma_{\mathrm{FRET}} \propto \Gamma_{0}\times z_0^{-4}$ since the dipole-dipole transfer rate scales quadratically with the transition dipole moment (with deviations from this scaling law at very small distances) \cite{gaudreau_universal_2013, malic_forster-induced_2014,koppens_graphene_2011, gomez-santos_fluorescence_2011}. The weak FRET effect observed in experiments is thus consistent with the low radiative decay rate ($\Gamma_{0}\sim 10^{5}\,\mathrm{s^{-1}}$) calculated in \cite{ivady_ab_2020, reimers_photoluminescence_2020} and disagrees with the large values ($\Gamma_{0}\sim 10^{9}\,\mathrm{s^{-1}}$) reported in \cite{whitefield_magnetic_2025,lee_intrinsic_2025}

From the TRPL data on SiO$_2$/Si (Fig.~\ref{fig2}\textcolor{blue}{c}), fitting with a mono-exponential convoluted with the instrument response function (IRF) yields
\begin{equation}
\Gamma_{0} + \Gamma_{\mathrm{nr}} \approx 10^{9}\,\mathrm{s^{-1}}
\end{equation}
consistent with earlier lifetime studies \cite{gottscholl_initialization_2020, clua-provost_spin-dependent_2024}. This confirms $\Gamma_{0} \ll \Gamma_{\mathrm{nr}}$. Two consequences follow: (i) the low quantum yield explains the weak photon count rates typically observed from $V_{B}^{-}$ centers, and (ii) $\Gamma_{\mathrm{FRET}} \ll \Gamma_{\mathrm{nr}}$ for most emitters in a 5 nm flake, making the PL nearly independent of whether the flake lies on MLG or SiO$_2$/Si.  

To observe FRET more clearly, the hBN must be thinned so that the emitters are much closer to MLG. Figures~\ref{fig2}d--f show the optical image, the cw PL and the TRPL from a 2 nm flake. Here, PL intensity is reduced by a factor of three on MLG, accompanied by a noticeably shorter decay time, consistent with $\Gamma_{\mathrm{FRET}}$ becoming competitive with $\Gamma_{\mathrm{nr}}$. We can exclude purely optical effects (\textit{e.g.}, modified absorption or collection efficiency) as the cause, since these would not alter the decay time and would not produce the strong contrast observed between 2 nm and 5 nm flakes (see Fig. S5a for more details). 

\begin{figure}[h!]
\label{Fig3}
\includegraphics[width=0.85\linewidth]{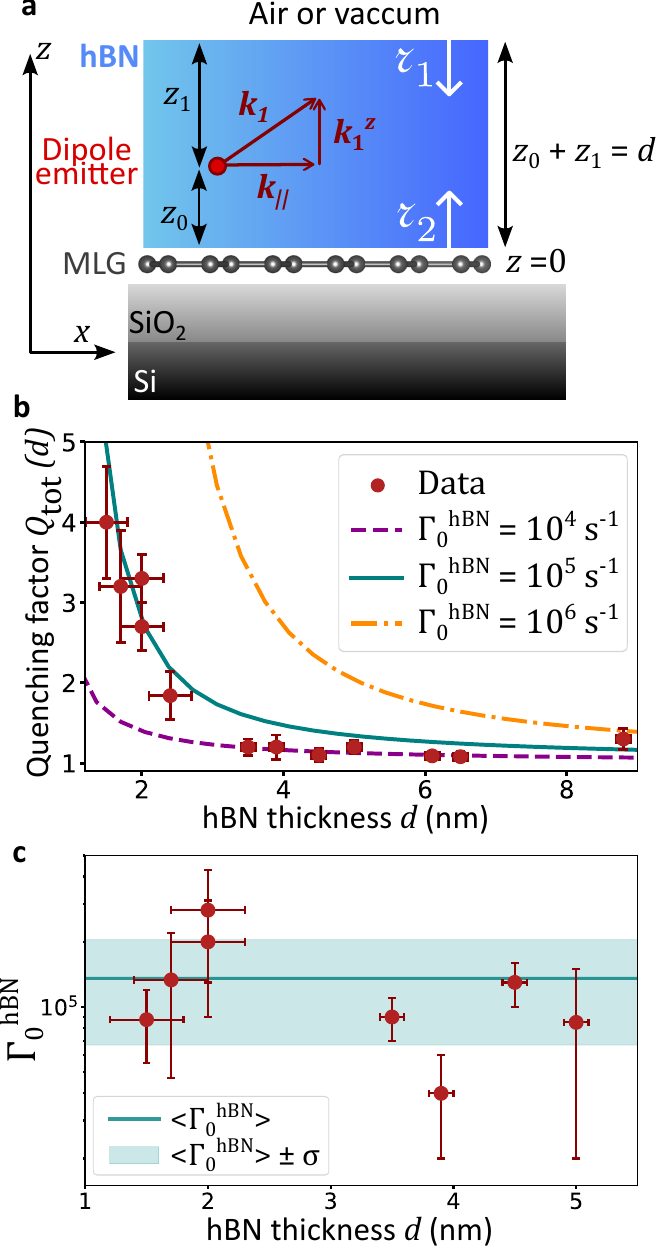}
\caption{a) Sketch of the structure used for calculating the decay rate. b) Quenching factor defined by Eq. \ref{Quenchdef} (data are shown in red points, results of the model for three values of $\Gamma^{\mathrm{hBN}}_{0}$ are shown by the green lines. c) $\Gamma^{\mathrm{hBN}}_{0}$ extracted from the TRPL data for each sample.}
\label{fig3}
\end{figure}

Figure~\ref{fig3}b summarizes the cw PL quenching factor, defined as
\begin{equation}
\label{Quenchdef}
Q_{\mathrm{tot}}(d) = \frac{I^{\mathrm{SiO}_{2}}_{\mathrm{tot}}(d)}{I^{\mathrm{MLG}}_{\mathrm{tot}}(d)}
\end{equation}
for flakes of thickness $d$, where $I^{\mathrm{SiO}_{2}}_{\mathrm{tot}}$ and $I^{\mathrm{MLG}}_{\mathrm{tot}}$ are the PL intensities detected on SiO$_2$ and MLG, respectively. Raster scans and TRPL data for each thickness are provided in the Supplemental Material \cite{supplementary-material}. Quenching becomes increasingly pronounced as the flakes become thinner, and is significant for $d < 3$ nm.  

We now show how an experimental value of the radiative rate can be extracted from $Q_{\mathrm{tot}}(d)$. This requires evaluating $\Gamma_{\mathrm{FRET}}$ for emitters at different distances $z_0$ from MLG. We assume a homogeneous defect density of independent emitters across the hBN layers, with identical FRET coupling for emitters within the same layer. In contrast to previous works where simplified model systems were studied including an emitter in vacuum above MLG on a semi-infinite substrate~\cite{gaudreau_universal_2013}, we need to take into account the fact that the reflection from both hBN interfaces, as well as from the SiO$_2$/Si interface modify the FRET. To that end we develop an electrodynamical model (see Supplemental Material \cite{supplementary-material}).

In brief, we consider a 0D emitter with in-plane dipole moment $\vect{d_{x}}$ radiating at frequency $\omega$. The emitted field decomposes into $s$- and $p$-polarized waves with the same in-plane wavevector component $k_\parallel$ in all layers and different absolute values $k_{i}$ and $z-$components $k_i^z$ in medium $i$ (0: air, 1: hBN, 2: SiO$_2$, 3: Si), where
\begin{equation}
k_{i} = \frac{\omega}{c}\sqrt{\varepsilon_{i}}, \quad 
k^{z}_{i} = \sqrt{k_{i}^{2}-k_{\parallel}^{2}}
\end{equation}
and $\varepsilon_i$ is the dielectric susceptibility of the medium $i$. We neglect optical anisotropy of hBN for simplicity.
The structure is illustrated in Fig.~\ref{fig3}a. The top (air/hBN) interface is described by the Fresnel coefficients $\mathscr r_1^\beta$, $\mathscr t_1^\beta$, while the bottom (hBN/graphene/SiO$_2$/Si) interface is described by an effective reflection coefficient $\mathscr r_2^\beta$ ($\beta = s, p$ denotes the polarization), obtained via the partial waves method including multiple reflections (details in Supplemental Material \cite{supplementary-material}). These multiple reflections enhance the emitted field by a factor $\mathscr E_\beta$ compared to an emitter in homogeneous hBN \cite{sipe_new_1987, tomas_green_1995}.
\begin{equation}
\label{Enh}
\begin{split}
   \mathscr E_\beta\left(z_0\right)  = &\frac{\left[1+\mathscr r_2^\beta \exp{(2\mathrm i z_0 k_{1}^{z})}\right]\left[1+\mathscr r_1^\beta \exp{(2\mathrm i z_1 k_{1}^{z})}\right] }{1-\mathscr r_1^\beta \mathscr r_2^\beta \exp{(2\mathrm i d k_{1}^{z})}}, \\\quad \beta = s, p 
\end{split}
\end{equation}
with $z_0$ and $z_1=d-z_0$ being the distances from the emitter to the bottom and top interfaces respectively.
The total decay rate of the emitter is then \cite{de_martini_spontaneous_1991, novotny_principles_2012, fang_control_2019}:
\begin{equation}
\label{Tot2}
   \Gamma_{\mathrm{tot}}\left(z_0\right) = \Gamma_{\mathrm{nr}}+\Gamma^{\mathrm{hBN}}_{0} \frac{3}{4k_{1}^{2}} \int_0^\infty d k_\parallel k_\parallel \Re{\left\{\frac{k_{1}}{k_{1}^z} \mathscr E_s + \frac{k_{1}^z}{k_{1}} \mathscr E_p \right\}},
\end{equation}
where the radiative decay rate of the emitter in homogeneous and infinite hBN is $$\Gamma^{\mathrm{hBN}}_{0}=\frac{1}{3\pi\varepsilon_{0}}\left(\frac{\omega}{c}\right)^3\sqrt{\varepsilon_{1}}\frac{|\vect{d_x}|^2}{\hbar}.$$
The photons collected by our microscope objective with a numerical aperture NA are emitted with an effective decay rate:
\begin{multline}
\label{Col}
\Gamma_{\mathrm{collec}}\left(z_0\right) = \Gamma^{\mathrm{hBN}}_{0} \frac{3}{8\sqrt{\varepsilon_{1}}k_{1}^{2}}\times \\
\int_0^{k_{0}\times \mathrm{NA}} d k_\parallel k_\parallel \Re{\left\{\left(\frac{k_{1}}{k_{1}^{z}}\right)^{2}\frac{k_{0}^{z}}{k_{0}} \mathscr U_s + \frac{k_{0}^{z}}{k_{0}} \mathscr U_p \right\}},
\end{multline}

with
\begin{equation}
	\label{Enhcollec}
		\mathscr U_\beta\left(z_0\right)  = \left|\mathscr t_1^\beta \frac{\exp{(\mathrm i z_1 k_{1}^{z})}+\mathscr r_2^\beta \exp{(\mathrm i d k_{1}^{z}+i z_0 k_{1}^{z})}}{1-\mathscr r_1^\beta \mathscr r_2^\beta \exp{(2\mathrm i d k_{1}^{z})}}\right|^{2}.
\end{equation}


The total PL intensity is obtained by summing the contribution of each single hBN layer $n$ at a distance $z_0$ from the MLG:
\begin{equation}
    I_{\mathrm{tot}}^{\mathrm{MLG,SiO_2}}(d)\propto\sum\limits^{N(d)}_{n=1}\dfrac{\Gamma_{\mathrm{collec}}(t_{ML}\times n)}{\Gamma_{\mathrm{tot}}(t_{ML}\times n)}
\end{equation}
where $n=1$ is the hBN layer in contact with graphene, $N(d) =  d/t_{ML}$ is the total number of hBN layers and $t_{ML}=0.34$ nm is the thickness of a hBN layer. The quenching factor is finally calculated using Eq. (\ref{Quenchdef}).

Results of the calculation are shown in Fig. \ref{fig3}b for three values of $\Gamma^{\mathrm{hBN}}_{0}$ (we fix $\Gamma_{\mathrm{nr}}=10^{9}$ s$^{-1}$). Very good agreement with our experimental data is obtained for $\Gamma^{\mathrm{hBN}}_{0}=10^{5}$ s$^{-1}$, consistent with ab-initio predictions \cite{ivady_ab_2020,reimers_photoluminescence_2020}.

The radiative decay rate of $V_B^-$, $\Gamma^{\mathrm{hBN}}_{0}$ can also be extracted from the TRPL data. The dynamics is fitted by a convolution of the IRF and a sum of monoexponential decays related to the emission of each layer $n$ at a distance $z_0$ from the MLG:
\begin{equation}
\label{TRPL}
    I_{\mathrm{tot}}^{\mathrm{MLG}}(d,t) \propto \sum\limits^{N(d)}_{n=1} \mathrm{exp}\left[ -t \times \Gamma_{\mathrm{tot}}(t_{ML}\times n)\right]
\end{equation}
The non-radiative decay rate $\Gamma_{\mathrm{nr}}$ was extracted for each sample by fitting the PL decay of hBN flakes on SiO$_2$ without graphene. As expected, $\Gamma_{\mathrm{nr}}$ is nearly independent of the hBN thickness (see Supplemental Material \cite{supplementary-material}). The decay dynamics on graphene were then fitted using Eq.~(\ref{TRPL}) for each sample, with $\Gamma_{0}^{\mathrm{hBN}}$ as the only free parameter (see Fig. \ref{fig2}c-f). The results are summarized in Fig. \ref{fig3}c, yielding
\[
\Gamma_{0}^{\mathrm{hBN}} = (1.35\pm0.68)\times10^{5}~\mathrm{s}^{-1}
\]

%% file: Sections/05-discussion.tex
Our results demonstrate that the PL intensity of thin hBN layers with $V_{B}^{-}$ centers is essentially insensitive to FRET for thicknesses greater than 3 nm. This insensitivity arises from the intrinsically low quantum yield of $V_{B}^{-}$ centers, since $\Gamma_{0}^{\mathrm{hBN}} \ll \Gamma_{\mathrm{nr}}$.
By contrast, emitters with a higher quantum yield are expected to be strongly affected by FRET, which could be a drawback for quantum sensing applications at very short distances. We show in the Supplemental Material \cite{supplementary-material} that a hBN flake of 8~nm with emitters with an intrinsic quantum yield of 1 would be quenched by a factor 100 when in contact with a MLG.
\begin{figure}[h!]
\label{Fig4}
\includegraphics[width=0.7\linewidth]{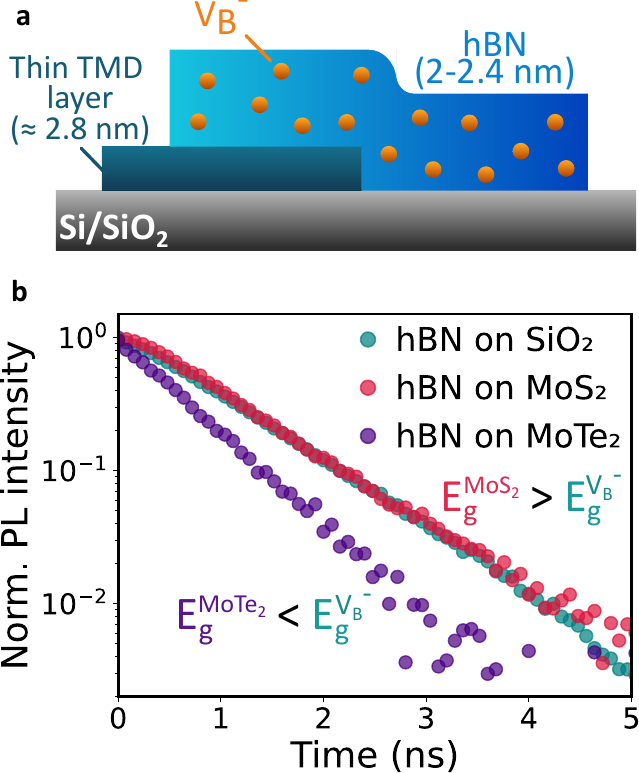}
\caption{a) Sketch of the sample to study FRET between $V_{B}^{-}$ centers and TMDs. b) TRPL data of a 2 nm hBN flake on top of SiO$_2$/Si, MoS$_2$ and MoTe$_2$. $\mathrm{E_g^{MoS_2}}$, $\mathrm{E_g^{MoS_2}}$ and $\mathrm{E_g^{V_B^-}}$ are the band gap energies of MoS$_2$, MoTe$_2$ and the emission energy of the $V_{B}^{-}$ centers.}
\label{fig4}
\end{figure}

A second advantage of $V_{B}^{-}$ centers stems from their low emission energy ($\sim 1.5$~eV). As a result, FRET is entirely suppressed when thin hBN layers are placed in contact with conventional 2D semiconductors having band gaps above 1.5~eV. To investigate this effect, we studied thin hBN flakes with $V_{B}^{-}$ centers coupled to two transition metal dichalcogenides (TMDs), MoS$_2$ and MoTe$_2$ (see Fig.~\ref{fig4}a). The PL decay dynamics for hBN on SiO$_2$, MoS$_2$, and MoTe$_2$ are shown in Fig.~\ref{fig4}b. Only for MoTe$_2$ is the lifetime clearly reduced, in agreement with the band gaps of the two TMDs (MoS$_2$: $\sim 2$ eV; MoTe$_2$: $\sim 1.1$~eV). This confirms that FRET occurs only when the semiconductor band gap lies below the emission energy of $V_{B}^{-}$. By contrast, most other color centers in hBN emit in the blue or green spectral range and are therefore expected to undergo strong FRET quenching in contact with nearly all TMDs.

%% file: Sections/06-conclusion.tex
In conclusion, we investigated FRET-mediated non-radiative energy transfer between $V_{B}^{-}$ centers in hBN and either MLG or 2D semiconductors. FRET is negligible for flakes thicker than 3 nm and entirely absent when the adjacent semiconductor has a band gap larger than 1.5 eV. Our results show that $V_{B}^{-}$ centers are well suited for integration into ultra-thin hBN sensing layers within van der Waals heterostructures. Furthermore, by extracting
\[
\Gamma_{0}^{\mathrm{hBN}} \sim 10^{5}\,\mathrm{s^{-1}},
\]
we unambiguously confirm that the weak emission of $V_{B}^{-}$ centers arises from their intrinsically low quantum yield. Our method to extract the radiative rate is highly versatile and can be extended to other defects in 2D materials or molecules grafted onto 2D layers, provided the emitter-graphene distance is known.

%% file: Sections/07-acknowledgements.tex
\\ \\
\textit{Aknowledgements:}
We thank S. Berciaud, A. Reserbat-Plantey, J-M. Gérard and N. Yao for fruitful discussions. This work was supported by Agence Nationale de la Recherche funding under the program ESR/EquipEx+ (grant number ANR-21-ESRE-0025), ANR-21-CE07-0021 (GRAAL), ANR QFoil, the grant NanoX n$^\circ$ ANR-17-EURE-0009 in the framework of the "Programme des Investissements d’Avenir", the Institute for Quantum Technologies in Occitanie through the project 2D-QSens and QuantEdu France. JHE and TP acknowledge the support from the National Science Foundation, award number 2413808, for hBN crystal growth. Neutron irradiation of the hBN crystals was supported by the U.S. Department of Energy, Office of Nuclear Energy under DOE Idaho Operations Office Contract DE-AC07-051D13417 as part of a Nuclear Science User Facilities experiment. The sample has been fabricated using the Exfolab platform.

%% file: SectionSupp/01_Sample_fabrication.tex
\newpage
\section{Sample fabrication}

Monolayer graphene was exfoliated from highly ordered pyrolytic graphite using the standard scotch tape method onto Si substrates with a 280 nm SiO$_{2}$ layer. The flakes were identified by their optical contrast under a microscope. Separately, hBN flakes were exfoliated onto Si substrates with a 80 nm SiO$_{2}$ layer to enhance visibility under white-light illumination. Thin flakes with thicknesses ranging from 1.5 to 9 nm were then transferred using the polycarbonate pick-up technique \cite{zomer_fast_2014} and positioned such that they partially covered both MLG and SiO$_{2}$(280 nm)/Si regions. Flake thicknesses were measured by atomic force microscopy (AFM).

The hBN bulk crystals used in this work were synthesized by metal flux growth, isotopically purified in $^{10}$B and $^{15}$N \cite{liu_single_2018}, and irradiated with neutrons at a fluence of 1.4 $\times$ 10$^{17}$ cm$^{-2}$ to create a homogeneous distribution of boron vacancies \cite{haykal_decoherence_2022}.

%% file: SectionSupp/02_Experimental_details.tex
\section{Experimental details}

Optical characterization was performed at room temperature using continuous-wave (cw) and time-resolved photoluminescence (TRPL). For cw PL, a 532 nm laser was focused onto the sample through a high-numerical-aperture objective (NA = 0.82), and the PL signal was collected by a single-photon avalanche detector (SPAD). Raster scans were acquired using a steering mirror. The excitation power was fixed at 400 µW.

For TRPL measurements, we employed a supercontinuum laser delivering 5 ps pulses (20 ps jitter) at a repetition rate of 40 MHz. A 510 nm excitation wavelength was selected using a laser line tunable filter (LLTF), with an average excitation power of 280 µW. The luminescence signal was recorded in photon-counting mode with a fiber-coupled SPAD (50 ps resolution) connected to a time-tagger with 80 ps binning, which set the effective time resolution of the experiment. The instrument response function (IRF) was obtained by measuring backscattered laser light from the substrate.

%% file: SectionSupp/03_Additional_data.tex
\newpage
\section{Additional data}
\begin{figure}[h!]
\includegraphics[width=\linewidth]{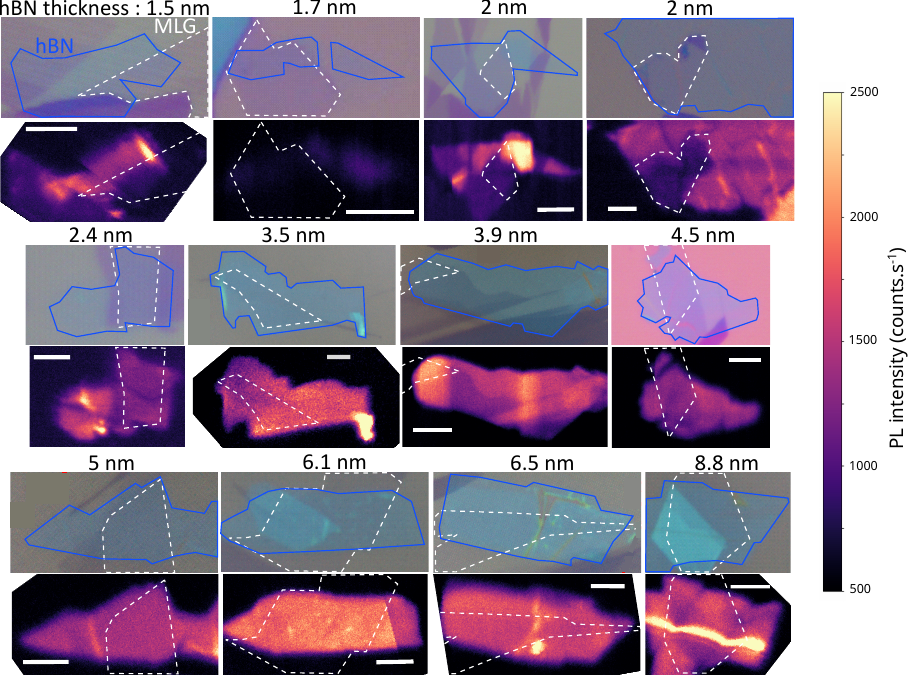}
\caption{Optical images and cw-PL raster scans for all samples. The scale bars is 5 µm.}
\label{supp_raster}
\end{figure}

\begin{figure}[h!]
\includegraphics[width=\linewidth]{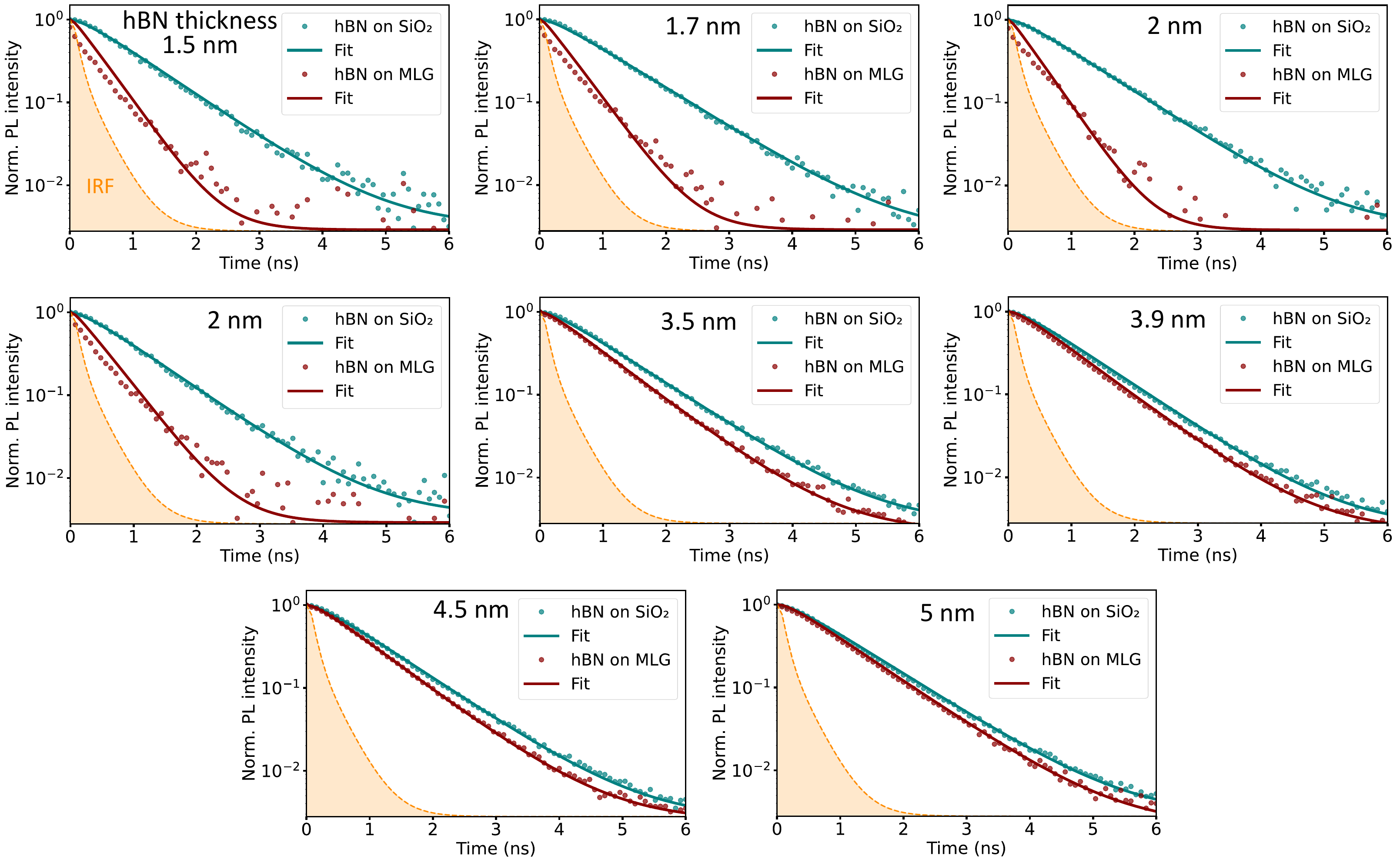}
\caption{TRPL data and fits for the samples with various thicknesses.}
\label{supp_trpl}
\end{figure}

\begin{figure}[h!]
\begin{center}
    \includegraphics[width=0.5\linewidth]{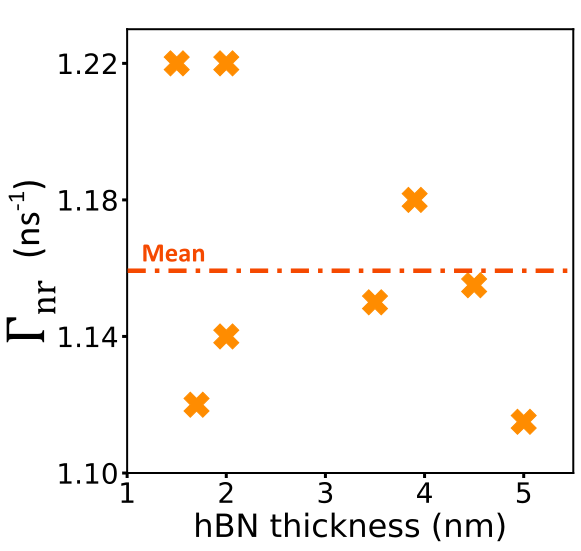}
    \caption{Relaxation rate from the excited state to the metastable state $\Gamma_\mathrm{nr}$ for each sample extracted from the fit of the TRPL data on SiO$_2$ (without graphene) using a monoexponential decay.}
    \label{supp_Gnr}
\end{center}
\end{figure}

Figure \ref{supp_raster} displays the PL raster scans for each sample, corresponding to various hBN thicknesses. Figure \ref{supp_trpl} presents the TRPL data along with the associated fits for each sample. Figure \ref{supp_Gnr} shows the values of $\Gamma_{\mathrm{nr}}$, extracted from the TRPL data, for each hBN flake on SiO$_{2}$. This analysis confirms that in the samples without graphene, $\Gamma_{\mathrm{nr}}$ remains largely independent of the hBN thickness.

%% file: SectionSupp/04_Model_test.tex
\newpage
\section{Electrodynamical model of Förster energy transfer}

\subsection{General model}

\begin{figure}[h]
\includegraphics[width=0.9\textwidth]{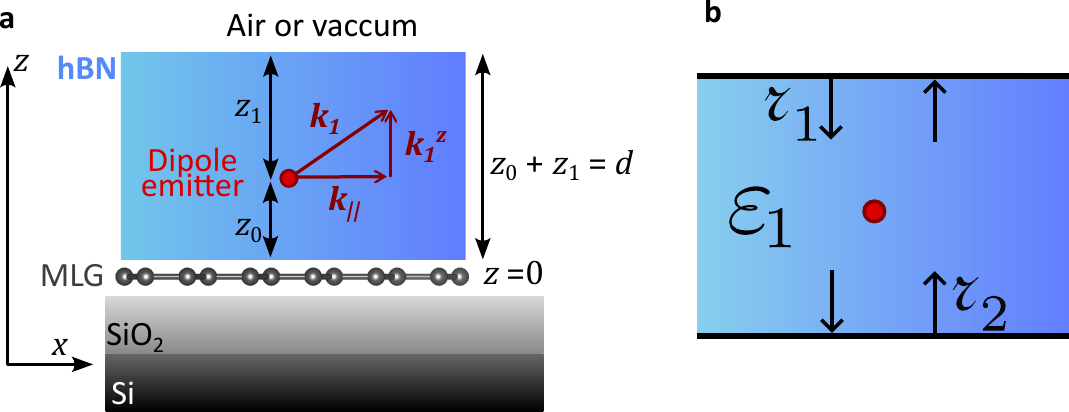}
\caption{(a) System under study. The system is assumed to be infinite in the $(xy)\perp z$ plane. The thickness of the hBN layer is $d$, the distance between the graphene layer and the emitter is $z_0$. The thickness of graphene is neglected. (b) Basic system consisting of the medium with the dielectric constant $\varepsilon$ surrounded with two (effective) mirrors with reflectivities $\mathscr r_1$ (top) and $\mathscr r_2$ (bottom).}\label{fig:scheme}
\end{figure}

We consider the system sketched in Fig.~\ref{fig:scheme}(a).
Let $\mathcal P_\alpha\delta(\bm r_0)$ be the Cartesian component of the polarization produced by the point emitter ($V_B^-$ center) positioned at the point $\bm r_0$. It satisfies the equation of motion~\cite{ivcenko_optical_2005}
\begin{equation}
\label{dipole}
\mathrm i \frac{d\mathcal P_\alpha}{dt} = (\omega_0 - \mathrm i \Gamma_{\rm nr})\mathcal P_\alpha - \frac{|d_{x}|^2}{\hbar}E_\alpha(\bm r_0),
\end{equation}
where $\omega_0$ is the emission frequency, $d_{x}$ is the effective transition-dipole matrix element, $E_\alpha$ is the component of electric field at the position of the emitter ($\alpha,\beta=x$, $y$ or $z$ denote corresponding Cartesian components), and $\Gamma_{\rm nr}$ includes all losses unrelated to the coupling with electromagnetic field (non-radiative losses). The electric field can be expressed via the polarization as~\cite{ivcenko_optical_2005, glazov_purcell_2011}
\begin{equation}
\label{field}
E_\alpha(\bm r) = 4\pi \left(\frac{\omega_0}{c}\right)^2 \sum_{\beta} G_{\alpha\beta}(\bm r, \bm r_0) \mathcal P_\beta,
\end{equation}
where we introduced the electromagnetic Green's function and neglected a minor difference between the bare emission frequency $\omega_0$ and the frequency renormalized by the light-matter interaction (Lamb shift).  Substituting Eq.~\eqref{field} into Eq.~\eqref{dipole} we obtain the decay rate of an electric dipole caused by its coupling with electromagnetic field in the medium as~\cite{ivcenko_optical_2005, glazov_purcell_2011, gomez-santos_fluorescence_2011}
\begin{equation}
\label{Gamma:0}
\Gamma = 4\pi \left(\frac{\omega_0}{c}\right)^2 \frac{|d_{x}|^2}{\hbar}\Im G_{\alpha\alpha}(\bm r_0,\bm r_0).
\end{equation}
Note that the decay rate in Eq.~\eqref{Gamma:0} includes both the radiative decay rate into propagating modes of the field in the medium and the decay related to the absorption in the medium, including FRET, see Ref.~\cite{glazov_purcell_2011} for details.

In what follows we focus on the experimentally relevant case where the microscopic dipole is oriented in the $xy$ plane. The case of an out-of-plane dipole will be presented in a separate work.

First, we derive the expression for the radiative decay rate of a point emitter for the general ``cavity-like'' system depicted in Fig.~\ref{fig:scheme}(b) as functions of the reflectivities $\mathscr r_1$ and $\mathscr r_2$ of the top and bottom interfaces. Afterwards we derive the expressions for $\mathscr r_1$ and $\mathscr r_2$ for the relevant system in Fig.~\ref{fig:scheme}(a). 

To obtain the general expression for the radiative decay rate, we use the method developed in Ref.~\cite{sipe_new_1987} to build the Green's function of the electromagnetic field. In the axially-symmetric system under study, it is convenient to separate all modes of the field into $s-$ and $p-$polarized ones. Consider a source which generates the electromagnetic wave of $\beta=s$- or $p$-polarization that propagates with the wavevector $\bm k_\parallel$ and frequency $\omega$ along the interfaces. Because of the multiple reflections the unit wave emitted by a source in the homogeneous medium (where, by definition, $\mathscr r_1=\mathscr r_2=0$) is enhanced by the factor
\begin{equation}
\label{Enhsupp}
   \mathscr E_\beta  = \frac{\left[1+\mathscr r_2^\beta \exp{(2\mathrm i z_0 k_{1}^{z})}\right]\left[1+\mathscr r_1^\beta \exp{(2\mathrm i z_1 k_{1}^{z})}\right] }{1-\mathscr r_1^\beta \mathscr r_2^\beta \exp{(2\mathrm i d k_{1}^{z})}}, \quad \beta = s, p 
\end{equation}
where $k_1$ is the amplitude of the wavevector in hBN that is decomposed between the parallel component $k_\parallel$ and the out-of-plane component $k_{1}^{z}$ (see Fig.~\ref{fig:scheme}):
\begin{equation}
\label{kz}
k_{1}^{z} = \sqrt{k_1^2 - k_\parallel^2}, \quad \mbox{with} \quad k_1 = \frac{\omega}{c}\sqrt{\varepsilon_1},
\end{equation}
$\varepsilon_1$ is the hBN dielectric susceptibility and we disregard the optical anisotropy of hBN for simplicity.
Note that $k_{1}^{z}$ can be, in general, complex where the imaginary part is related to evanescent waves. Here we use a convention where the reflection coefficients of $s$- and $p$-polarizations at a normal incidence have the same sign.

\subsection{Decay rate}
The decay rate~\eqref{Gamma:0} related to the coupling with electromagnetic field can be recast in the following form [cf. Ref.~\cite{de_martini_spontaneous_1991, novotny_principles_2012, koppens_graphene_2011, fang_control_2019}]
\begin{equation}\label{rad}
    \Gamma = \frac{\omega}{4\pi \varepsilon_0 c\sqrt{\varepsilon_1}}\frac{|d_x|^2}{\hbar}\int_0^\infty d k_\parallel k_\parallel \Re{\left\{\frac{k_1}{k_{1}^{z}} \mathscr E_s + \frac{k_{1}^{z}}{k_1} \mathscr E_p \right\}},
\end{equation}
where $d_x$, as before, is the transition dipole.

For emission into an infinite and homogeneous hBN media we have $\mathscr E_\beta=1$ and~\cite{beresteckij_quantum_2008, goupalov_light_2003, novotny_principles_2012} 
\begin{equation}
\label{Gamma:hom}
\Gamma_{0}^{\rm hBN} = \frac{1}{3\pi\varepsilon_0}\left(\frac{\omega}{c}\right)^3\sqrt{\varepsilon_1}\frac{|d_x|^2}{\hbar},
\end{equation}
where we made use of the following integrals
\[
\int_0^{k_1} d k_\parallel k_\parallel \frac{k_1}{k_{1}^{z}} = k_1^2, \quad \int_0^{k_1} d k_\parallel k_\parallel \frac{k_{1}^{z}}{ k_1}=k_1^2/3.
\]
Note that in the case of the emission in homogeneous hBN, the states with $k_\parallel>k_1$ play no role. Technically, it is because at $k_\parallel > k_1$ the $z$-component of the wavevector $k_1^z$ becomes purely imaginary and, in homogeneous medium where $\mathcal E_s = \mathcal E_p=1$, the real part of the subintegral expression vanishes. Physically, in homogeneous dielectric system there are no waves propagating with $k_\parallel> k_1$.

We can rewrite Eq.~\eqref{rad} as:
\begin{equation}
	\label{Tot3}
	\Gamma = \Gamma^{\mathrm{hBN}}_{0} \frac{3}{4k_{1}^{2}} \int_0^\infty d k_\parallel k_\parallel \Re{\left\{\frac{k_{1}}{k_{1}^z} \mathscr E_s + \frac{k_{1}^z}{k_{1}} \mathscr E_p \right\}},
\end{equation}
which corresponds to Eq. (7) of the main text.

The rate of the light emitted in the upward direction is calculated using the enhancement factor:
\begin{equation}
	\label{Enhcollecsupp}
	\mathscr U_\beta\left(z_0\right)  = \left|\mathscr t_1^\beta \frac{\exp{(\mathrm i z_1 k_{1}^{z})}+\mathscr r_2^\beta \exp{(\mathrm i d k_{1}^{z}+i z_0 k_{1}^{z})}}{1-\mathscr r_1^\beta \mathscr r_2^\beta \exp{(2\mathrm i d k_{1}^{z})}}\right|^{2},
\end{equation}
with $\mathscr t_1$ being the Fresnel transmission coefficient of the hBN/air top interface.

The portion of the light that is detected by a microscope objective with the numerical aperture NA is calculated by limiting the integration between $k_\parallel=0$ and $k_0\times \mathrm{NA}$ (with $k_0=\omega/c$) and considering only the light propagating upwards:
\begin{equation}
	\label{Colsupp}
	\Gamma_{\mathrm{collec}}\left(z_0\right) = \Gamma^{\mathrm{hBN}}_{0} \frac{3}{8\sqrt{\varepsilon_{1}}k_{1}^{2}}\times
	\int_0^{k_{0}\times \mathrm{NA}} d k_\parallel k_\parallel \Re{\left\{\left(\frac{k_{1}}{k_{1}^{z}}\right)^{2}\frac{k_{0}^{z}}{k_{0}} \mathscr U_s + \frac{k_{0}^{z}}{k_{0}} \mathscr U_p \right\}},
\end{equation}

\subsection{Effective reflective coefficients $\mathscr r_1$ and $\mathscr r_2$ for the structure hBN/graphene/SiO$_2$/Si}
We come back to the structure sketched in Fig.~\ref{fig:scheme}(a) and calculate the effective reflection coefficients $\mathscr r_1$ and $\mathscr r_2$ that enter the expression for the enhancement factor~\eqref{Enhsupp}. We define the index $i$ of each material such as $i=0$ for air, $i=1$ for hBN, $i=2$ for SiO$_2$ and $i=3$ for Si.
The Fresnel coefficients for light incident from the layer $i$ to $i+1$ are written as:
\begin{equation}
\label{rsp}
r_{i,i+1}^{s} = \frac{k_{i}^z - k_{i+1}^z}{k_{i}^z + k_{i+1}^z}
\end{equation}
\[
r_{i,i+1}^{p} = - \frac{n_{i+1}^2 k_{i}^z - n_{i}^2 k_{i+1}^z}{n_{j+1}^2 k_{i}^z + n_{j}^2 k_{i+1}^z}
\]
The transmission coefficients read
\begin{equation}
\label{tsp}
t_{i,i+1}^{s}=1+r_{i,i+1}^{s}
\end{equation}
\[
{t_{i,i+1}^{p}=\frac{n_{i}}{n_{i+1}} \left(1-r_{i,i+1}^{p}\right)}
\]
where $n_{i}$ is the complex refractive index of medium $i$ and $k_{i}^z$ is defined as:
\begin{equation}
k_{i}^{z} = \sqrt{k_i^2 - k_\parallel^2}, \quad \mbox{with} \quad k_i = \frac{\omega}{c}n_{i}.
\end{equation}

Now we use the partial wave summation technique to calculate the reflection coefficients of each interface.

\subsubsection{Interface hBN-air}
For our structure, we have simply:
\begin{equation}
\label{rr1}
\mathscr r_1^{s} =  r_{1,0}^{s} \quad \mbox{and} \quad \mathscr r_1^{p} =  r_{1,0}^{p}
\end{equation}

\subsubsection{Interface hBN/graphene/SiO$_2$/Si}
For the bottom interface, we first define the transmission and reflection coefficients of a monolayer graphene in vacuum. 
\begin{equation}
\label{tgr}
t^{s}_{gr}= \frac{1}{1+ \frac{\sigma}{2 \varepsilon_0 c} \frac{k_{0}}{k_{0}^z}} , \quad t^{p}_{gr}= \frac{1}{1+ \frac{\sigma}{2 \varepsilon_0 c} \frac{k_{0}^z}{k_{0}}}.
\end{equation}
\begin{equation}
\label{rgr}
r_{gr}^{s} = t^{s}_{gr} - 1 , \quad r_{gr}^{p} = t^{p}_{gr} - 1.
\end{equation}
$\sigma=\frac{e^{2}}{4\hbar}$ is the sheet conductivity of graphene. We assume that the Fermi level in graphene deposited on the SiO$_{2}$ surface is very small (a few 10's of meV; \textit{i.e.} the doping is small) as compared to energy of the emitted photons. We can neglect the plasmonic effects \cite{koppens_graphene_2011}. We also neglect the effect of temperature \cite{koppens_graphene_2011}, and disregard the effects of space dispersion on graphene's conductivity leaving it for the future work.
Absorbance at normal incidence is $\mathcal A = 1-|t_{gr}|^2 -|r_{gr}|^2 \approx \frac{\sigma}{\varepsilon_0 c}=\pi \alpha$, where $\alpha=\frac{e^2}{4\pi \varepsilon_0 \hbar c}$ is the fine structure constant.

Then we define the effective reflection coefficient of the structure vacuum/SiO$_2$/Si:
\begin{equation}
\label{r023}
    r_{0,2,3}^{\beta}=r_{0,2}^{\beta} + \frac{t_{0,2}^{\beta}t_{2,0}^{\beta}r_{2,3}^{\beta} \exp \left(2 \mathrm i k_{2}^z L_2\right)}{1-r_{2,3}^{\beta}r_{2,0}^{\beta}\exp \left(2 \mathrm i k_{2}^z L_2\right)}
\end{equation}
where $L_2$ is the SiO$_2$ thickness.

The effective reflection coefficient of the structure vacuum/graphene/SiO$_2$/Si writes:
\begin{equation}
\label{rbg}
    r_{bg}^{\beta}=r_{gr}^{\beta} + \frac{\left(t_{gr}^{\beta}\right)^{2}r_{0,2,3}^{\beta}}{1-r_{02,3}^{\beta}r_{gr}^{\beta}}
\end{equation}

Finally, the effective reflection coefficient for the structure hBN/graphene/SiO$_2$/Si is:
\begin{equation}
\label{r2}
    \mathscr r_2^{\beta}=r_{1,0}^{\beta} + \frac{t_{1,0}^{\beta}t_{0,1}^{\beta}r_{bg}^{\beta}}{1-r_{0,1}^{\beta}r_{bg}^{\beta}}
\end{equation}

\subsection{Calculation of $\Gamma_{\mathrm{collec}}$ and $\Gamma$}
\begin{figure}[h!]
\begin{center}
\includegraphics[width=1\linewidth]{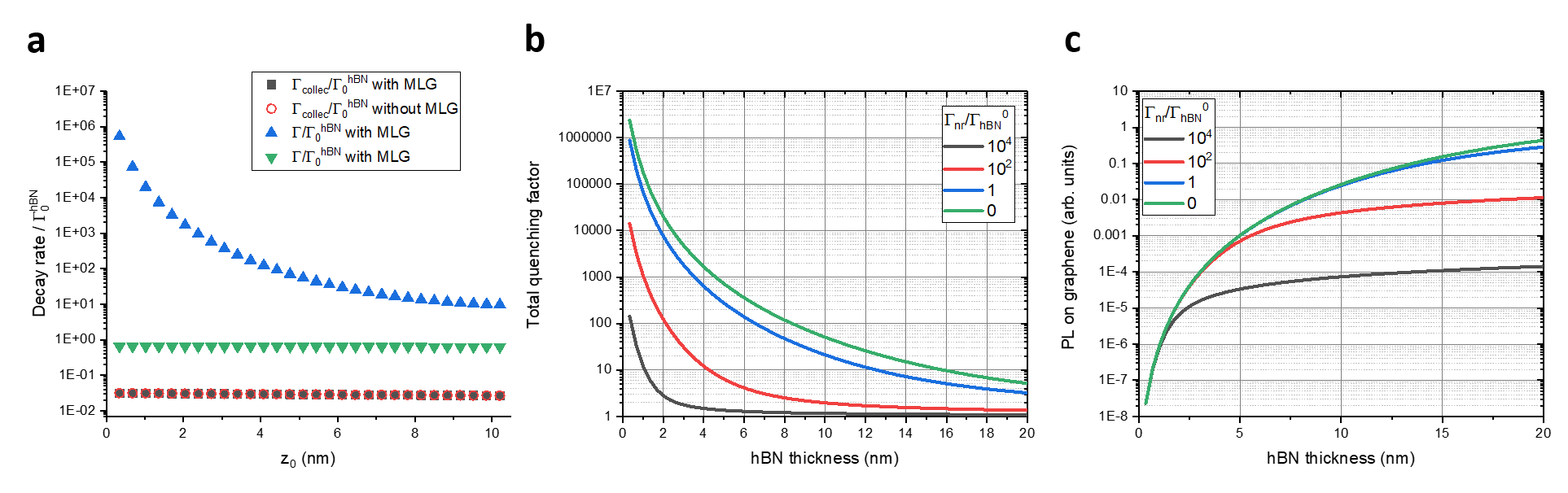}
    \caption{a) Ratio of $\frac{\Gamma_{\mathrm{collec}}}{\Gamma_0^{\mathrm{hBN}}}$ and $\frac{\Gamma}{\Gamma_0^{\mathrm{hBN}}}$ for the structure on SiO$_2$/Si with and without graphene and on a platinum substrate. b) Total quenching factor defined as the total PL intensity on SiO$_2$/Si divided by the total PL intensity on MLG/SiO$_2$/Si. c) Total PL of a hBN flake on graphene.}
    \label{supp_Gcollec}
\end{center}
\end{figure}

Fig.\ref{supp_Gcollec}(a) presents the results of our calculation of $\Gamma_{\mathrm{collec}}$ and $\Gamma$ as function of the position of the emitter $z_0$ for a 30 layers thick (10.2 nm) hBN flake on top of MLG/SiO$_{2(280\text{nm})}$/Si and SiO$_{2(280\text{nm})}$/Si. The refractive indexes used are summarized in Table \ref{table_index}.

\begin{table}[h!]
\centering
\begin{tabular}{|c|c|c|c|c|}
\hline
\rowcolor[HTML]{EFEFEF} 
Material                                 & air ($i=0$) & hBN ($i=1$) & SiO$_2$ ($i=2$) & Si ($i=3$) \\ \hline
\cellcolor[HTML]{EFEFEF}Refractive index & 1         & 2.2       & 1.46       & 3.5      \\ \hline
\end{tabular}
\caption{Refractive indexes used for the calculations.}
\label{table_index}
\end{table}

The results show that $\Gamma_{\mathrm{collec}}$ is the same with or without MLG (black and red points) which justify our assumption of neglecting the change of collection efficiency on the measured PL quenching. Without graphene, the ratio $\frac{\Gamma_{\mathrm{collec}}}{\Gamma}$ is around 15$\%$ which means that a significant fraction of the light is not collected by our microscope objective.

The blue points show the increasing of the FRET effect at small $z_0$. Nevertheless, we see that for $z_0>1.2$~nm, $\frac{\Gamma}{\Gamma^{\mathrm{hBN}}_0}<10^4$, which means that for $V_{B}^{-}$ centers with $\Gamma^{\mathrm{hBN}}_0 \approx10^5~\mathrm s^{-1}$, $\Gamma<\Gamma_{\mathrm{nr}}=10^9~\mathrm{s}^{-1}$.

For emitters with larger intrinsic quantum yield (\textit{i.e.} smaller $\frac{\Gamma_{\mathrm{nr}}}{\Gamma^{\mathrm{hBN}}_0}$), the FRET is more competitive at larger $z_0$. In Fig.\ref{supp_Gcollec}(b), we present the total PL quenching factor defined by Eq. (4) of the main text as a function of the hBN thickness and for various ratio $\frac{\Gamma_{\mathrm{nr}}}{\Gamma^{\mathrm{hBN}}_0}$. We see that for an emitter with an intrinsic quantum yield of 1 ($\frac{\Gamma_{\mathrm{nr}}}{\Gamma^{\mathrm{hBN}}_0}=0$), the quenching factor is as high as 100 for a 8~nm thick hBN flake.

It is also instructive to plot the variation of the total PL intensity for a hBN flake on graphene as a function of the ratio $\frac{\Gamma_{\mathrm{nr}}}{\Gamma^{\mathrm{hBN}}_0}$. Results presented in Fig.\ref{supp_Gcollec}(c) show that, for a 3 nm flake, the total PL intensity of a $V_{B}^{-}$ centers is only six times lower than that of a perfect emitter. For a 1 nm flake, the difference between $V_{B}^{-}$ and a perfect emitter becomes negligible.
These findings demonstrate that the quantum yield of the emitter is not a critical factor in enhancing the sensitivity of ultra-thin quantum sensing layers in contact with graphene.

%% file: references.bib
@article{steiner_current-induced_2025,
	title = {Current-induced brightening of vacancy-related emitters in hexagonal boron nitride},
	volume = {7},
	issn = {2643-1564},
	url = {https://link.aps.org/doi/10.1103/cd62-5hq8},
	doi = {10.1103/cd62-5hq8},
	number = {3},
	urldate = {2025-09-09},
	journal = {Physical Review Research},
	author = {Steiner, Corinne and Rahmel, Rebecca and Volmer, Frank and Windisch, Rika and Janssen, Lars H. and Pesch, Patricia and Watanabe, Kenji and Taniguchi, Takashi and Libisch, Florian and Beschoten, Bernd and Stampfer, Christoph and Kurzmann, Annika},
	month = aug,
	year = {2025},
	pages = {L032037},
}

@article{healey_quantum_2023,
	title = {Quantum microscopy with van der {Waals} heterostructures},
	volume = {19},
	issn = {1745-2473, 1745-2481},
	url = {https://www.nature.com/articles/s41567-022-01815-5},
	doi = {10.1038/s41567-022-01815-5},
	number = {1},
	urldate = {2025-09-09},
	journal = {Nature Physics},
	author = {Healey, A. J. and Scholten, S. C. and Yang, T. and Scott, J. A. and Abrahams, G. J. and Robertson, I. O. and Hou, X. F. and Guo, Y. F. and Rahman, S. and Lu, Y. and Kianinia, M. and Aharonovich, I. and Tetienne, J.-P.},
	month = jan,
	year = {2023},
	pages = {87--91},
}

@article{durand_optically_2023,
	title = {Optically active spin defects in few-layer thick hexagonal boron nitride},
	volume = {131},
	issn = {0031-9007, 1079-7114},
	url = {https://link.aps.org/doi/10.1103/PhysRevLett.131.116902},
	doi = {10.1103/PhysRevLett.131.116902},
	number = {11},
	urldate = {2025-09-09},
	journal = {Physical Review Letters},
	author = {Durand, A. and Clua-Provost, T. and Fabre, F. and Kumar, P. and Li, J. and Edgar, J.H. and Udvarhelyi, P. and Gali, A. and Marie, X. and Robert, C. and Gérard, J-M. and Gil, B. and Cassabois, G. and Jacques, V.},
	month = sep,
	year = {2023},
	pages = {116902},
}

@article{fang_control_2019,
	title = {Control of the exciton radiative lifetime in van der {Waals} heterostructures},
	volume = {123},
	issn = {0031-9007, 1079-7114},
	url = {https://link.aps.org/doi/10.1103/PhysRevLett.123.067401},
	doi = {10.1103/physrevlett.123.067401},
	number = {6},
	urldate = {2025-07-26},
	journal = {Physical Review Letters},
	author = {Fang, H.H. and Han, B. and Robert, C. and Semina, M.A. and Lagarde, D. and Courtade, E. and Taniguchi, T. and Watanabe, K. and Amand, T. and Urbaszek, B. and Glazov, M.M. and Marie, X.},
	month = aug,
	year = {2019},
	pages = {067401}
}

@book{novotny_principles_2012,
	address = {Cambridge},
	edition = {2nd ed},
	title = {Principles of {Nano}-{Optics}},
	isbn = {978-1-107-00546-4 978-1-139-55425-1},
	publisher = {Cambridge University Press},
	author = {Novotny, Lukas},
	collaborator = {Hecht, Bert},
	year = {2012},
}

@article{de_martini_spontaneous_1991,
	title = {Spontaneous emission in the optical microscopic cavity},
	volume = {43},
	copyright = {http://link.aps.org/licenses/aps-default-license},
	issn = {1050-2947, 1094-1622},
	url = {https://link.aps.org/doi/10.1103/PhysRevA.43.2480},
	doi = {10.1103/physreva.43.2480},
	number = {5},
	urldate = {2025-07-26},
	journal = {Physical Review A},
	author = {De Martini, F. and Marrocco, M. and Mataloni, P. and Crescentini, L. and Loudon, R.},
	month = mar,
	year = {1991},
	pages = {2480--2497},
}

@article{tomas_green_1995,
	title = {Green function for multilayers: {Light} scattering in planar cavities},
	volume = {51},
	copyright = {http://link.aps.org/licenses/aps-default-license},
	issn = {1050-2947, 1094-1622},
	shorttitle = {Green function for multilayers},
	url = {https://link.aps.org/doi/10.1103/PhysRevA.51.2545},
	doi = {10.1103/PhysRevA.51.2545},
	number = {3},
	urldate = {2025-09-09},
	journal = {Physical Review A},
	author = {Tomaš, M. S.},
	month = mar,
	year = {1995},
	pages = {2545--2559},
}

@article{sipe_new_1987,
	title = {New {Green}-function formalism for surface optics},
	volume = {4},
	copyright = {https://doi.org/10.1364/OA\_License\_v1\#VOR},
	issn = {0740-3224, 1520-8540},
	url = {https://opg.optica.org/abstract.cfm?URI=josab-4-4-481},
	doi = {10.1364/JOSAB.4.000481},
	number = {4},
	urldate = {2025-07-29},
	journal = {Journal of the Optical Society of America B},
	author = {Sipe, J. E.},
	month = apr,
	year = {1987},
	pages = {481},
}

@article{gomez-santos_fluorescence_2011,
	title = {Fluorescence quenching in graphene: {A} fundamental ruler and evidence for transverse plasmons},
	volume = {84},
	copyright = {http://link.aps.org/licenses/aps-default-license},
	issn = {1098-0121, 1550-235X},
	shorttitle = {Fluorescence quenching in graphene},
	url = {https://link.aps.org/doi/10.1103/PhysRevB.84.165438},
	doi = {10.1103/PhysRevB.84.165438},
	number = {16},
	urldate = {2025-09-09},
	journal = {Physical Review B},
	author = {Gómez-Santos, G. and Stauber, T.},
	month = oct,
	year = {2011},
	pages = {165438},
}

@article{clua-provost_spin-dependent_2024,
	title = {Spin-dependent photodynamics of boron-vacancy centers in hexagonal boron nitride},
	volume = {110},
	issn = {2469-9950, 2469-9969},
	url = {https://link.aps.org/doi/10.1103/PhysRevB.110.014104},
	doi = {10.1103/PhysRevB.110.014104},
	number = {1},
	urldate = {2025-09-09},
	journal = {Physical Review B},
	author = {Clua-Provost, T. and Mu, Z. and Durand, A. and Schrader, C. and Happacher, J. and Bocquel, J. and Maletinsky, P. and Fraunié, J. and Marie, X. and Robert, C. and Seine, G. and Janzen, E. and Edgar, J. H. and Gil, B. and Cassabois, G. and Jacques, V.},
	month = jul,
	year = {2024},
	pages = {014104},
}

@article{ivady_ab_2020,
	title = {Ab initio theory of the negatively charged boron vacancy qubit in hexagonal boron nitride},
	volume = {6},
	issn = {2057-3960},
	url = {https://www.nature.com/articles/s41524-020-0305-x},
	doi = {10.1038/s41524-020-0305-x},
	number = {1},
	urldate = {2023-02-23},
	journal = {npj Computational Materials},
	author = {Ivády, Viktor and Barcza, Gergely and Thiering, Gergő and Li, Song and Hamdi, Hanen and Chou, Jyh-Pin and Legeza, Örs and Gali, Adam},
	month = apr,
	year = {2020},
	pages = {41},
}

@article{reimers_photoluminescence_2020,
	title = {Photoluminescence, photophysics, and photochemistry of the {$V_B^-$} defect in hexagonal boron nitride},
	volume = {102},
	issn = {2469-9950, 2469-9969},
	url = {https://link.aps.org/doi/10.1103/PhysRevB.102.144105},
	doi = {10.1103/PhysRevB.102.144105},
	number = {14},
	urldate = {2022-10-11},
	journal = {Physical Review B},
	author = {Reimers, Jeffrey R. and Shen, Jun and Kianinia, Mehran and Bradac, Carlo and Aharonovich, Igor and Ford, Michael J. and Piecuch, Piotr},
	month = oct,
	year = {2020},
	pages = {144105},
}

@article{haykal_decoherence_2022,
	title = {Decoherence of {$V_B^-$} spin defects in monoisotopic hexagonal boron nitride},
	volume = {13},
	issn = {2041-1723},
	doi = {10.1038/s41467-022-31743-0},
	number = {1},
	urldate = {2022-10-10},
	journal = {Nature Communications},
	author = {Haykal, A. and Tanos, R. and Minotto, N. and Durand, A. and Fabre, F. and Li, J. and Edgar, J. H. and Ivady, V. and Gali, A. and Michel, T. and Dréau, A. and Gil, B. and Cassabois, G. and Jacques, V.},
	year = {2022},
	pages = {4347},
}

@article{liu_single_2018,
	title = {Single {Crystal} {Growth} of {Millimeter}-{Sized} {Monoisotopic} {Hexagonal} {Boron} {Nitride}},
	volume = {30},
	issn = {0897-4756, 1520-5002},
	url = {https://pubs.acs.org/doi/10.1021/acs.chemmater.8b02589},
	doi = {10.1021/acs.chemmater.8b02589},
	number = {18},
	urldate = {2025-01-28},
	journal = {Chemistry of Materials},
	author = {Liu, Song and He, Rui and Xue, Lianjie and Li, Jiahan and Liu, Bin and Edgar, James H.},
	month = sep,
	year = {2018},
	pages = {6222--6225},
}

@article{zomer_fast_2014,
	title = {Fast pick up technique for high quality heterostructures of bilayer graphene and hexagonal boron nitride},
	volume = {105},
	issn = {0003-6951, 1077-3118},
	url = {https://pubs.aip.org/apl/article/105/1/013101/596305/Fast-pick-up-technique-for-high-quality},
	doi = {10.1063/1.4886096},
	number = {1},
	urldate = {2025-09-09},
	journal = {Applied Physics Letters},
	author = {Zomer, P. J. and Guimarães, M. H. D. and Brant, J. C. and Tombros, N. and Van Wees, B. J.},
	month = jul,
	year = {2014},
	pages = {013101},
}

@article{vaidya_quantum_2023,
	title = {Quantum sensing and imaging with spin defects in hexagonal boron nitride},
	volume = {8},
	issn = {2374-6149},
	url = {https://www.tandfonline.com/doi/full/10.1080/23746149.2023.2206049},
	doi = {10.1080/23746149.2023.2206049},
	number = {1},
	urldate = {2025-09-09},
	journal = {Advances in Physics: X},
	author = {Vaidya, Sumukh and Gao, Xingyu and Dikshit, Saakshi and Aharonovich, Igor and Li, Tongcang},
	month = dec,
	year = {2023},
	pages = {2206049},
}

@article{tisler_single_2013,
	title = {Single defect center scanning near-field optical microscopy on graphene},
	volume = {13},
	issn = {1530-6984, 1530-6992},
	url = {https://pubs.acs.org/doi/10.1021/nl401129m},
	doi = {10.1021/nl401129m},
	number = {7},
	urldate = {2024-04-15},
	journal = {Nano Letters},
	author = {Tisler, Julia and Oeckinghaus, Thomas and Stöhr, Rainer J. and Kolesov, Roman and Reuter, Rolf and Reinhard, Friedemann and Wrachtrup, Jörg},
	month = jul,
	year = {2013},
	pages = {3152--3156},
}

@article{chen_energy_2010,
	title = {Energy transfer from individual semiconductor nanocrystals to graphene},
	volume = {4},
	issn = {1936-0851, 1936-086X},
	url = {https://pubs.acs.org/doi/10.1021/nn1005107},
	doi = {10.1021/nn1005107},
	number = {5},
	urldate = {2025-07-26},
	journal = {ACS Nano},
	author = {Chen, Zheyuan and Berciaud, Stéphane and Nuckolls, Colin and Heinz, Tony F. and Brus, Louis E.},
	month = may,
	year = {2010},
	pages = {2964--2968},
}

@article{koppens_graphene_2011,
	title = {Graphene plasmonics: a platform for strong light–matter interactions},
	volume = {11},
	issn = {1530-6984, 1530-6992},
	shorttitle = {Graphene {Plasmonics}},
	url = {https://pubs.acs.org/doi/10.1021/nl201771h},
	doi = {10.1021/nl201771h},
	number = {8},
	urldate = {2025-07-26},
	journal = {Nano Letters},
	author = {Koppens, Frank H. L. and Chang, Darrick E. and García De Abajo, F. Javier},
	month = aug,
	year = {2011},
	pages = {3370--3377},
}

@article{malic_forster-induced_2014,
	title = {Förster-induced energy transfer in functionalized graphene},
	volume = {118},
	copyright = {http://pubs.acs.org/page/policy/authorchoice\_ccby\_termsofuse.html},
	issn = {1932-7447, 1932-7455},
	url = {https://pubs.acs.org/doi/10.1021/jp5019636},
	doi = {10.1021/jp5019636},
	number = {17},
	urldate = {2025-07-26},
	journal = {The Journal of Physical Chemistry C},
	author = {Malic, Ermin and Appel, Heiko and Hofmann, Oliver T. and Rubio, Angel},
	month = may,
	year = {2014},
	pages = {9283--9289},
}

@article{federspiel_distance_2015,
	title = {Distance dependence of the energy transfer rate from a single semiconductor nanostructure to graphene},
	volume = {15},
	issn = {1530-6984, 1530-6992},
	url = {https://pubs.acs.org/doi/10.1021/nl5044192},
	doi = {10.1021/nl5044192},
	number = {2},
	urldate = {2025-07-26},
	journal = {Nano Letters},
	author = {Federspiel, François and Froehlicher, Guillaume and Nasilowski, Michel and Pedetti, Silvia and Mahmood, Ather and Doudin, Bernard and Park, Serin and Lee, Jeong-O and Halley, David and Dubertret, Benoît and Gilliot, Pierre and Berciaud, Stéphane},
	month = feb,
	year = {2015},
	pages = {1252--1258},
}

@article{gaudreau_universal_2013,
	title = {Universal distance-scaling of nonradiative energy transfer to graphene},
	volume = {13},
	issn = {1530-6984, 1530-6992},
	url = {https://pubs.acs.org/doi/10.1021/nl400176b},
	doi = {10.1021/nl400176b},
	number = {5},
	urldate = {2024-07-04},
	journal = {Nano Letters},
	author = {Gaudreau, L. and Tielrooij, K. J. and Prawiroatmodjo, G. E. D. K. and Osmond, J. and De Abajo, F. J. García and Koppens, F. H. L.},
	month = may,
	year = {2013},
	pages = {2030--2035},
}

@article{dexter_theory_1953,
	title = {A theory of sensitized luminescence in solids},
	volume = {21},
	issn = {0021-9606, 1089-7690},
	url = {https://pubs.aip.org/jcp/article/21/5/836/203559/A-Theory-of-Sensitized-Luminescence-in-Solids},
	doi = {10.1063/1.1699044},
	number = {5},
	urldate = {2025-06-23},
	journal = {The Journal of Chemical Physics},
	author = {Dexter, D. L.},
	month = may,
	year = {1953},
	pages = {836--850},
}

@article{forster_zwischenmolekulare_1948,
	title = {Zwischenmolekulare {Energiewanderung} und {Fluoreszenz}},
	volume = {437},
	issn = {0003-3804, 1521-3889},
	url = {https://onlinelibrary.wiley.com/doi/10.1002/andp.19484370105},
	doi = {10.1002/andp.19484370105},
	number = {1-2},
	urldate = {2025-06-23},
	journal = {Annalen der Physik},
	author = {Förster, Th.},
	month = jan,
	year = {1948},
	pages = {55--75},
}

@article{fraunie_charge_2025,
	title = {Charge state tuning of spin defects in hexagonal boron nitride},
	volume = {25},
	copyright = {https://doi.org/10.15223/policy-029},
	issn = {1530-6984, 1530-6992},
	url = {https://pubs.acs.org/doi/10.1021/acs.nanolett.5c00654},
	doi = {10.1021/acs.nanolett.5c00654},
	number = {14},
	urldate = {2025-09-09},
	journal = {Nano Letters},
	author = {Fraunié, J. and Clua-Provost, T. and Roux, S. and Mu, Z. and Delpoux, A. and Seine, G. and Lagarde, D. and Watanabe, K. and Taniguchi, T. and Marie, X. and Poirier, T. and Edgar, J. H. and Grisolia, J. and Lassagne, B. and Claverie, A. and Jacques, V. and Robert, C.},
	month = apr,
	year = {2025},
	pages = {5836--5842},
}

@article{zhou_sensing_2024,
	title = {Sensing spin wave excitations by spin defects in few-layer-thick hexagonal boron nitride},
	volume = {10},
	issn = {2375-2548},
	url = {https://www.science.org/doi/10.1126/sciadv.adk8495},
	doi = {10.1126/sciadv.adk8495},
	number = {18},
	urldate = {2024-07-04},
	journal = {Science Advances},
	author = {Zhou, Jingcheng and Lu, Hanyi and Chen, Di and Huang, Mengqi and Yan, Gerald Q. and Al-matouq, Faris and Chang, Jiu and Djugba, Dziga and Jiang, Zhigang and Wang, Hailong and Du, Chunhui Rita},
	month = may,
	year = {2024},
	pages = {eadk8495},
}

@article{huang_wide_2022,
	title = {Wide field imaging of van der {Waals} ferromagnet {Fe$_3$GeTe$_2$} by spin defects in hexagonal boron nitride},
	volume = {13},
	issn = {2041-1723},
	url = {https://www.nature.com/articles/s41467-022-33016-2},
	doi = {10.1038/s41467-022-33016-2},
	number = {1},
	urldate = {2022-10-10},
	journal = {Nature Communications},
	author = {Huang, Mengqi and Zhou, Jingcheng and Chen, Di and Lu, Hanyi and McLaughlin, Nathan J. and Li, Senlei and Alghamdi, Mohammed and Djugba, Dziga and Shi, Jing and Wang, Hailong and Du, Chunhui Rita},
	month = sep,
	year = {2022},
	pages = {5369},
}

@article{kumar_pawan_magnetic_2022,
	title = {Magnetic imaging with spin defects in hexagonal boron nitride},
	volume = {18},
	url = {https://link.aps.org/doi/10.1103/PhysRevApplied.18.L061002},
	doi = {10.1103/PhysRevApplied.18.L061002},
	number = {6},
	urldate = {2025-04-10},
	journal = {Physical Review Applied},
	author = {Kumar, P. and Fabre, F. and Durand, A. and Clua-Provost, T. and Li, J. and Edgar, J.H. and Rougemaille, N. and Coraux, J. and Marie, X. and Renucci, P. and Robert, C. and Robert-Philip, I. and Gil, B. and Cassabois, G. and Finco, A. and Jacques, V.},
	month = dec,
	year = {2022},
	pages = {061002},
}

@article{gottscholl_spin_2021,
	title = {Spin defects in hBN as promising temperature, pressure and magnetic field quantum sensors},
	volume = {12},
	issn = {2041-1723},
	url = {https://www.nature.com/articles/s41467-021-24725-1},
	doi = {10.1038/s41467-021-24725-1},
	number = {1},
	urldate = {2022-10-10},
	journal = {Nature Communications},
	author = {Gottscholl, Andreas and Diez, Matthias and Soltamov, Victor and Kasper, Christian and Krauße, Dominik and Sperlich, Andreas and Kianinia, Mehran and Bradac, Carlo and Aharonovich, Igor and Dyakonov, Vladimir},
	month = dec,
	year = {2021},
	pages = {4480},
}

@article{gottscholl_initialization_2020,
	title = {Initialization and read-out of intrinsic spin defects in a van der {Waals} crystal at room temperature},
	volume = {19},
	issn = {1476-1122, 1476-4660},
	url = {https://www.nature.com/articles/s41563-020-0619-6},
	doi = {10.1038/s41563-020-0619-6},
	number = {5},
	urldate = {2022-10-25},
	journal = {Nature Materials},
	author = {Gottscholl, Andreas and Kianinia, Mehran and Soltamov, Victor and Orlinskii, Sergei and Mamin, Georgy and Bradac, Carlo and Kasper, Christian and Krambrock, Klaus and Sperlich, Andreas and Toth, Milos and Aharonovich, Igor and Dyakonov, Vladimir},
	month = may,
	year = {2020},
	pages = {540--545},
}

@article{aharonovich_quantum_2022,
	title = {Quantum emitters in hexagonal boron nitride},
	volume = {22},
	issn = {1530-6984, 1530-6992},
	url = {https://pubs.acs.org/doi/10.1021/acs.nanolett.2c03743},
	doi = {10.1021/acs.nanolett.2c03743},
	urldate = {2022-11-28},
	journal = {Nano Letters},
	author = {Aharonovich, Igor and Tetienne, Jean-Philippe and Toth, Milos},
	year = {2022},
    pages = {9227},
}

@article{white_electrical_2022,
	title = {Electrical control of quantum emitters in a van der {Waals} heterostructure},
	volume = {11},
	issn = {2047-7538},
	url = {https://www.nature.com/articles/s41377-022-00877-7},
	doi = {10.1038/s41377-022-00877-7},
	number = {1},
	urldate = {2024-04-04},
	journal = {Light: Science \& Applications},
	author = {White, Simon J. U. and Yang, Tieshan and Dontschuk, Nikolai and Li, Chi and Xu, Zai-Quan and Kianinia, Mehran and Stacey, Alastair and Toth, Milos and Aharonovich, Igor},
	month = jun,
	year = {2022},
	pages = {186},
}

@article{yu_electrical_2022,
	title = {Electrical charge control of h-{BN} single photon sources},
	volume = {9},
	issn = {2053-1583},
	url = {https://iopscience.iop.org/article/10.1088/2053-1583/ac75f4},
	doi = {10.1088/2053-1583/ac75f4},
	number = {3},
	urldate = {2024-04-04},
	journal = {2D Materials},
	author = {Yu, Mihyang and Yim, Donggyu and Seo, Hosung and Lee, Jieun},
	month = jul,
	year = {2022},
	pages = {035020},
}

@article{grotz_charge_2012,
	title = {Charge state manipulation of qubits in diamond},
	volume = {3},
	issn = {2041-1723},
	url = {https://www.nature.com/articles/ncomms1729},
	doi = {10.1038/ncomms1729},
	number = {1},
	urldate = {2024-04-15},
	journal = {Nature Communications},
	author = {Grotz, Bernhard and Hauf, Moritz V. and Dankerl, Markus and Naydenov, Boris and Pezzagna, Sébastien and Meijer, Jan and Jelezko, Fedor and Wrachtrup, Jörg and Stutzmann, Martin and Reinhard, Friedemann and Garrido, Jose A.},
	month = mar,
	year = {2012},
	pages = {729},
}

@article{hauf_chemical_2011,
	title = {Chemical control of the charge state of nitrogen-vacancy centers in diamond},
	volume = {83},
	copyright = {http://link.aps.org/licenses/aps-default-license},
	issn = {1098-0121, 1550-235X},
	url = {https://link.aps.org/doi/10.1103/PhysRevB.83.081304},
	doi = {10.1103/PhysRevB.83.081304},
	number = {8},
	urldate = {2025-04-28},
	journal = {Physical Review B},
	author = {Hauf, M. V. and Grotz, B. and Naydenov, B. and Dankerl, M. and Pezzagna, S. and Meijer, J. and Jelezko, F. and Wrachtrup, J. and Stutzmann, M. and Reinhard, F. and Garrido, J. A.},
	month = feb,
	year = {2011},
	pages = {081304},
}

@article{santori_vertical_2009,
	title = {Vertical distribution of nitrogen-vacancy centers in diamond formed by ion implantation and annealing},
	volume = {79},
	copyright = {http://link.aps.org/licenses/aps-default-license},
	issn = {1098-0121, 1550-235X},
	url = {https://link.aps.org/doi/10.1103/PhysRevB.79.125313},
	doi = {10.1103/PhysRevB.79.125313},
	number = {12},
	urldate = {2025-04-28},
	journal = {Physical Review B},
	author = {Santori, Charles and Barclay, Paul E. and Fu, Kai-Mei C. and Beausoleil, Raymond G.},
	month = mar,
	year = {2009},
	pages = {125313},
}

@article{gerard_quantum_2024,
	title = {Quantum efficiency and vertical position of quantum emitters in hBN determined by {Purcell} effect in hybrid metal-dielectric planar photonic structures},
	volume = {11},
	url = {https://doi.org/10.1021/acsphotonics.4c01416},
	doi = {10.1021/acsphotonics.4c01416},
	number = {12},
	urldate = {2025-09-30},
	journal = {ACS Photonics},
	author = {Gérard, Domitille and Pierret, AurÃ©lie and Fartas, Helmi and Berini, Bruno and Buil, StÃ©phanie and Hermier, Jean-Pierre and Delteil, Aymeric},
	month = dec,
	year = {2024},
	pages = {5188--5194},
}

@article{swathi_resonance_2008,
	title = {Resonance energy transfer from a dye molecule to graphene},
	volume = {129},
	issn = {0021-9606},
	url = {https://doi.org/10.1063/1.2956498},
	doi = {10.1063/1.2956498},
	number = {5},
	urldate = {2025-11-12},
	journal = {The Journal of Chemical Physics},
	author = {Swathi, R. S. and Sebastian, K. L.},
	month = aug,
	year = {2008},
	pages = {054703},
}

@article{basko_energy_2000,
	title = {Energy transfer from a semiconductor quantum dot to an organic matrix},
	volume = {13},
	issn = {1434-6036},
	url = {https://doi.org/10.1007/s100510050082},
	doi = {10.1007/s100510050082},
	number = {4},
	urldate = {2025-11-12},
	journal = {The European Physical Journal B - Condensed Matter and Complex Systems},
	author = {Basko, D.M. and Agranovich, V.M. and Bassani, F. and Rocca, G.C. La},
	month = feb,
	year = {2000},
	pages = {653--659},
}

@article{kos_different_2005,
	title = {Different regimes of {Förster}-type energy transfer between an epitaxial quantum well and a proximal monolayer of semiconductor nanocrystals},
	volume = {71},
	url = {https://link.aps.org/doi/10.1103/PhysRevB.71.205309},
	doi = {10.1103/PhysRevB.71.205309},
	number = {20},
	urldate = {2025-11-12},
	journal = {Physical Review B},
	author = {Kos, A. and Achermann, M. and Klimov, V. I. and Smith, D. L.},
	month = may,
	year = {2005},
	pages = {205309},
}

@article{lee_intrinsic_2025,
	title = {Intrinsic high-fidelity spin polarization of charged vacancies in hexagonal boron nitride},
	volume = {134},
	doi ={10.1103/PhysRevLett.134.096202},
	journal = {Physical Review Letters},
	author = {Lee, W. and Liu, V.S. and Zhang, Z. and Kim, S. and Gong, R. and Du, X. and Pham, K. and Poirier, T. and Hao, Z. and Edgar, J.H. and Kim, P. and Zu, C. and Davis, E.J. and Yao, N.A.},
	year = {2025},
	pages = {096202},
}

@article{whitefield_magnetic_2025,
	title = {Magnetic field sensitivity optimization of negatively charged boron vacancy defects in {hBN}.},
	volume = {8},
	doi = {10.1002/qute.202300118},
	journal = {Advanced Quantum Technologies},
	author = {Whitefield, Benjamin and Toth, Milos and Aharonovich, Igor and Tetienne, Jean-Philippe and Kianinia, Mehran},
	year = {2025},
	pages = {2300118},
}

@article{glazov_purcell_2011,
	title = {Purcell factor in small metallic cavities},
	volume = {53},
	copyright = {http://www.springer.com/tdm},
	issn = {1063-7834, 1090-6460},
	url = {http://link.springer.com/10.1134/S1063783411090125},
	doi = {10.1134/S1063783411090125},
	urldate = {2025-09-19},
	journal = {Physics of the Solid State},
	author = {Glazov, M. M. and Ivchenko, E. L. and Poddubny, A. N. and Khitrova, G.},
	year = {2011},
	pages = {1753--1760},
}

@book{ivcenko_optical_2005,
	address = {Harrow, UK},
	title = {Optical spectroscopy of semiconductor nanostructures},
	isbn = {978-1-84265-150-6},
	publisher = {Alpha Science International Ltd},
	author = {Ivchenko, E. L.},
	year = {2005},
}

@book{beresteckij_quantum_2008,
	address = {Oxford},
	edition = {2. ed., reprint},
	series = {Course of theoretical physics / {L}. {D}. {Landau} and {E}. {M}. {Lifshitz}},
	title = {Quantum electrodynamics},
	isbn = {978-0-7506-3371-0},
	number = {4},
	publisher = {Butterworth-Heinemann},
	author = {Berestetskii, V. B. and Lifshitz, E. M. and Pitaevskii, L. P.},
	year = {2008},
}

@article{goupalov_light_2003,
	title = {Light scattering on exciton resonance in a semiconductor quantum dot: {Exact} solution},
	volume = {68},
	copyright = {http://link.aps.org/licenses/aps-default-license},
	issn = {0163-1829, 1095-3795},
	shorttitle = {Light scattering on exciton resonance in a semiconductor quantum dot},
	url = {https://link.aps.org/doi/10.1103/PhysRevB.68.125311},
	doi = {10.1103/PhysRevB.68.125311},
	urldate = {2025-09-19},
	journal = {Physical Review B},
	author = {Goupalov, S. V.},
	year = {2003},
	pages = {125311},
}


%% file: references2.bib
@misc{supplementary-material,
	note = {See Supplemental Material for additional experimental details and derivation of the model which includes Refs. \cite{zomer_fast_2014,liu_single_2018,haykal_decoherence_2022,glazov_purcell_2011,ivcenko_optical_2005,beresteckij_quantum_2008,goupalov_light_2003}},
}
